# 1 Ω–10 kΩ high Precision transportable Setup to calibrate Multifunction Electrical instruments

P.P. Capra[1] and F. Galliana[2]

*National Institute of Metrological Research, (INRIM) str. Cacce, 91 – 10135 (TURIN Italy)*
*[1] Phone + 39 011 3919424, fax + 39 011 3919448, p.capra@inrim.it*
*[2] Phone + 39 011 3919336, fax + 39 011 3919448, f.galliana@inrim.it*

## ABSTRACT

A temperature controlled 1 Ω–10 kΩ standard Resistors transportable setup was developed at National Institute of Metrological Research, (INRIM) for the calibration and adjustment of multifunction electrical instruments. The two Standards consist respectively of two 10 Ω and 100 kΩ parallel connected resistors nets inserted in a temperature controlled aluminium box. Novelty of the realization is the oil insertion of the 1 Ω net with its internal connectors lowering the thermo-electromotive forces (emfs) effects. Short and mid-term stabilities of the setup Standards resulted on the order and in some cases better than other top level 1 Ω and 10 kΩ commercial Standards. The transport effect turning off the setup temperature control did not cause appreciable measurement deviations on the two Standards. The Standards uncertainties meet those requested by DMMs and MFCs manufacturers to calibrate and adjust these instruments. A test to adjust a multifunction calibrator gave satisfactory results.

**Key Words**: standard resistor, multifunction calibrator (MFC), digital multimeter (DMM), Resistance measurements, measurement stability, power and temperature coefficients, measurement uncertainties.

## 1. INTRODUCTION

The need to develop, maintain, compare and use for traceability transfer high accuracy 1 Ω and 10 kΩ Resistance Standards had been felt since some decades by National Metrological Institutes (NMI) [1–5]. High accuracy multifunction electrical instruments as digital multimeters (DMMs) and multifunction calibrators (MFCs) operating in particular in the five low frequency electrical quantities (DC and AC Voltage, DC and AC current and DC Resistance), widely used as Standards in calibration electrical laboratories can be calibrated by means of a process called "artifact calibration" that requires few reference Standards among which the 1 Ω and 10 kΩ Resistance Standards. This process allows to these instruments to self-assign new values to their internal references. [6–9]. To transport only few Standards for the calibration of DMM's and MFC's, instead of more and more delicate instruments or Standards increases the traceability transfer accuracy, reliability and convenience. For this reason, at National Institute of Metrological Research, (INRIM) a temperature controlled 1 Ω and 10 kΩ standard Resistors setup was developed to calibrate and adjust DMM's and MFC's. This setup could be also involved as local Standard to avoid thermal enclosures often necessary for high accuracy primary Resistance Standards [2] or specially made [10]. In addition the setup Standards could act as traveling Standards for international Comparisons (ILC's) as in [5] or in [11] (for other values) and for national ILC's as in [12]. The actual setup involving the two main Resistance values for the traceability transfer to DMM's and MFC's is an improvement and upgrading of a first attempt to develop a thermo-regulated standard Resistor made at INRIM with encouraging results [13]. Construction details, stability tests also in comparison with top level 1 Ω and 10 kΩ commercial Resistors, temperature and power coefficients, tests on the transport effect and of a MFC adjustment, evaluation of the use[1] uncertainties as local laboratory Standards and for calibration of electrical instruments for the setup 1 Ω–10 kΩ Standards are given.

## 2. THE SETUP STANDARDS NETWORKS

The setup involves two resistors nets with Vishay VHA 512 type resistors, with tolerance of ± 0.001 %, temperature coefficient (TCR) less than $2\times10^{-6}$/°C and long term stability of $5\times10^{-6}$/year according to the manufacturer specifications. For the 1 Ω standard Resistor ten matched 10 Ω resistors were connected in parallel with their leads and a manganin strap, chosen instead of copper due to its lower TCR. The 10 kΩ Standard was made by means of a net of ten 100 kΩ matched and parallel connected resistors. Its parallel connection was made with a manganin strap as for the 1 Ω net.

## 3. THE THERMOSTATIC BOX

The two resistors nets were inserted inside an aluminum box (Fig. 1). The 1 Ω resistors net was placed into a cylindrical space inside the box filled with mineral oil to enhance the resistors-box temperature exchange. The novelty was to put in oil this resistors net with its internal connectors to maintain the temperature uniformity among them to reduce the thermal emfs. The net was connected to four binding post external connectors on the box cover fixed with thermal conductive resin to uniform the temperature among the connectors. The 10 kΩ resistors net was placed into ten holes in an external ring of the box. The bottom of the box was mechanically connected to a Peltier element (thermoelectric cooling, TEC) connected to a radiator outside the box. The box was placed in a metallic case filled with polystyrene foam.

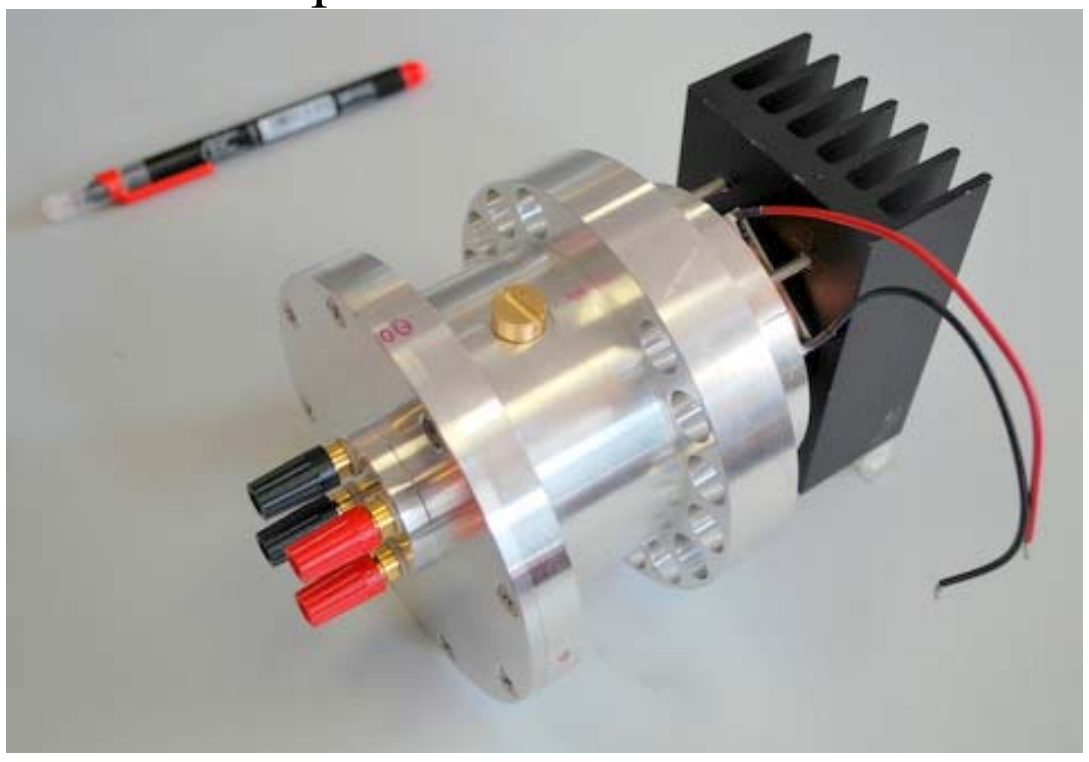

Fig. 1. View of the aluminium box connected to a TEC and a radiator.

The TEC is supplied by a Proportional-Integral-Derivative (PID) controller put in another case with the microcontroller and the power supply (Fig. 2).

### 3.1 Temperature control system

The temperature-control of the box is based on a commercial low noise PID controller with a Negative Temperature Coefficient temperature sensor (NTC). The system can operate in stand-alone or in pc-controlled mode. In stand-alone mode, the controller checks the box and environment temperatures, the status of the battery and the display.

[1] We define use uncertainty the effective uncertainty that a standard or instrument introduces in the time period between two its calibrations when it is used to calibrate other standards or instruments. It normally comprehends its calibration, drift, environment conditions and other influence parameters dependence uncertainty components. Similar description was given in [14].

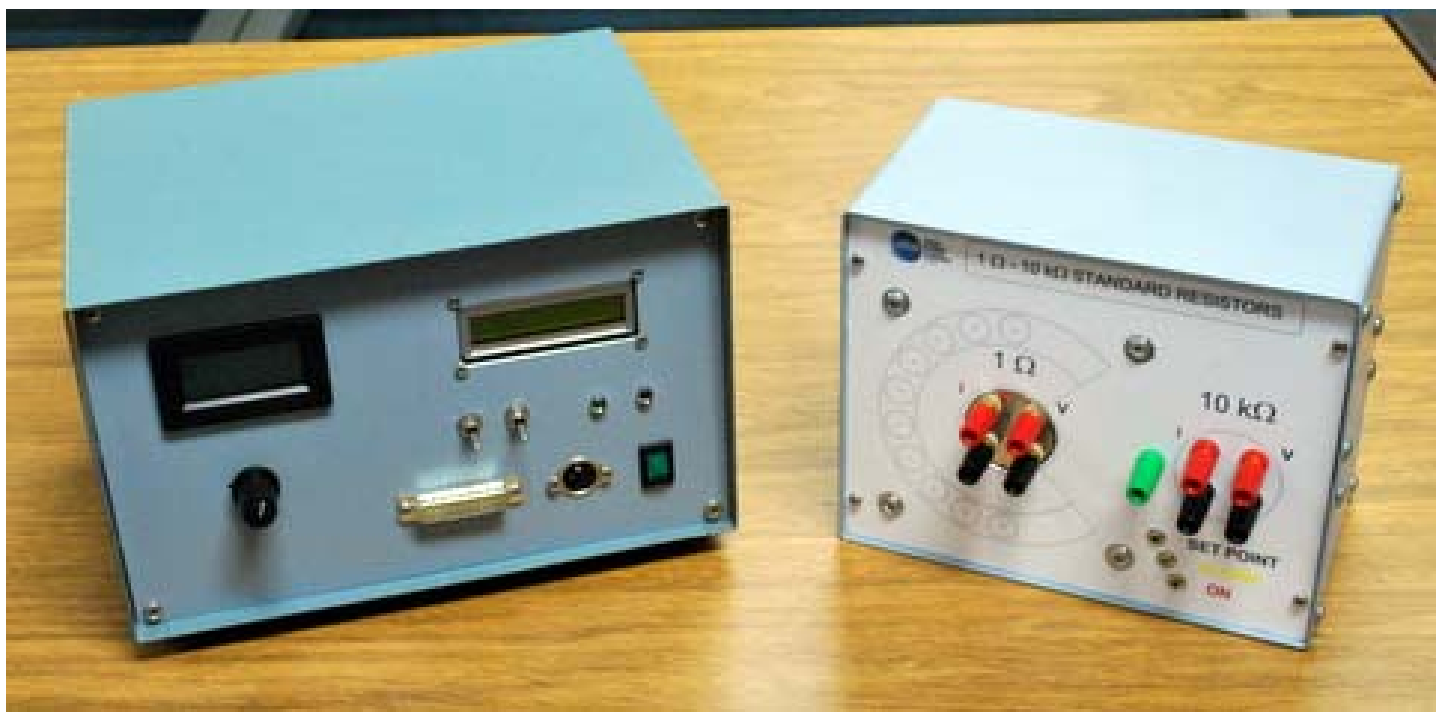

Fig. 2. Temperature controller system (left) and Standards boxes (right).

Fig. 3 shows the main frame of the program to read and set the temperature of the resistors box. When the controller operates in pc-controlled mode the display shows the temperature set point, the environment and box temperatures and the last calibration values of the Standards. By means of a USB-pc connection, it's possible to change the temperature set-point, load the box and laboratory temperatures and store the Standards calibration data on the microcontroller memory.

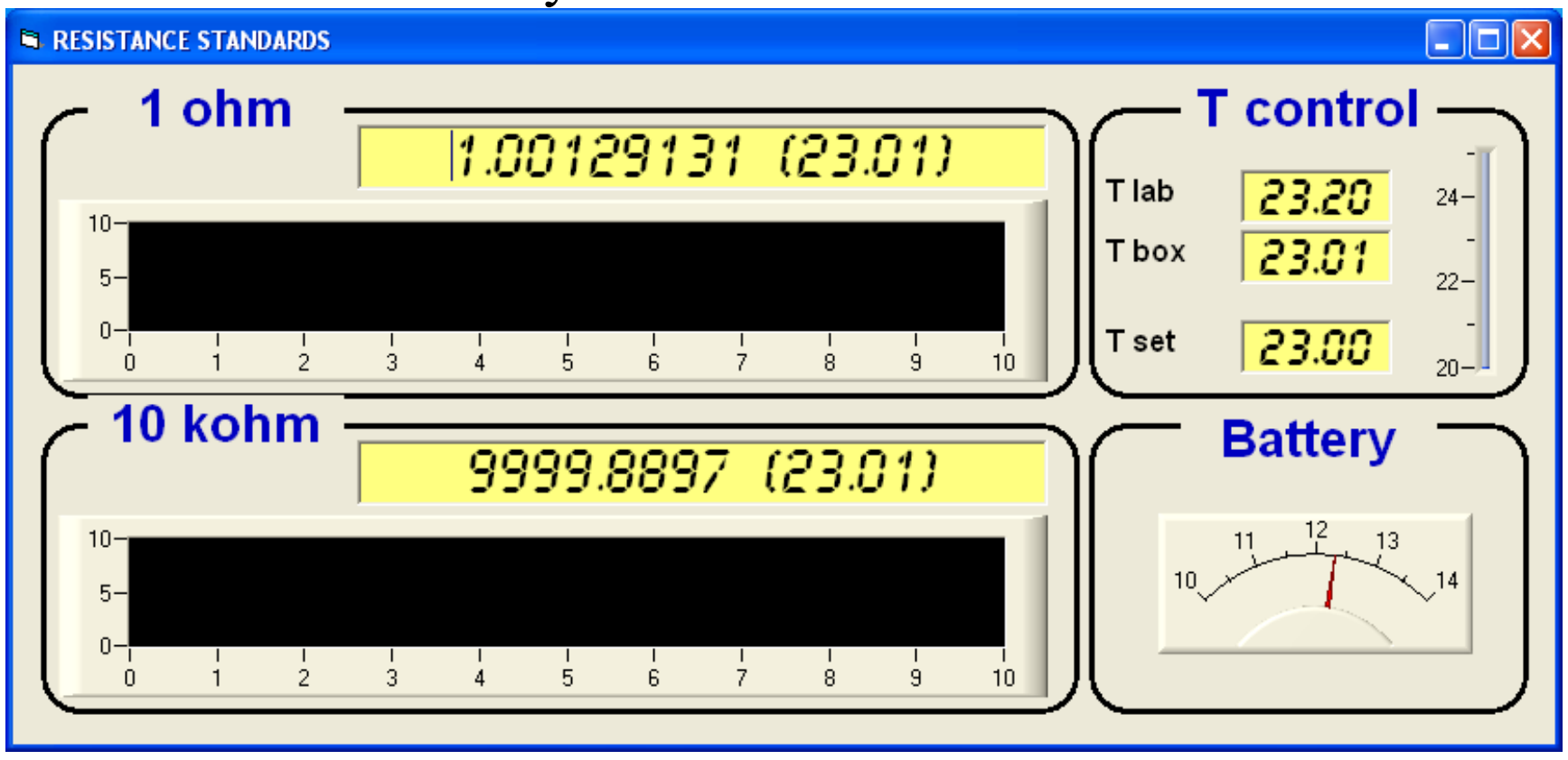

Fig. 3. Main frame of the control program of the setup.

The programs to control the parameters of the Standards and the firmware of the microcontroller were respectively written in Visual Basic and C.

### 3.2 Efficiency of the temperature control

Figure 4 shows the 2 h temperature stability of the box with the temperature controller set at 23 °C. After a transient due to the temperature set point change, the stability is better than 5 mK. The system needs about 30 min to change the temperature in a range of about 3 degrees around 23 °C to reach the desired stability in a thermo-regulated laboratory at (23 ± 0.5) °C.

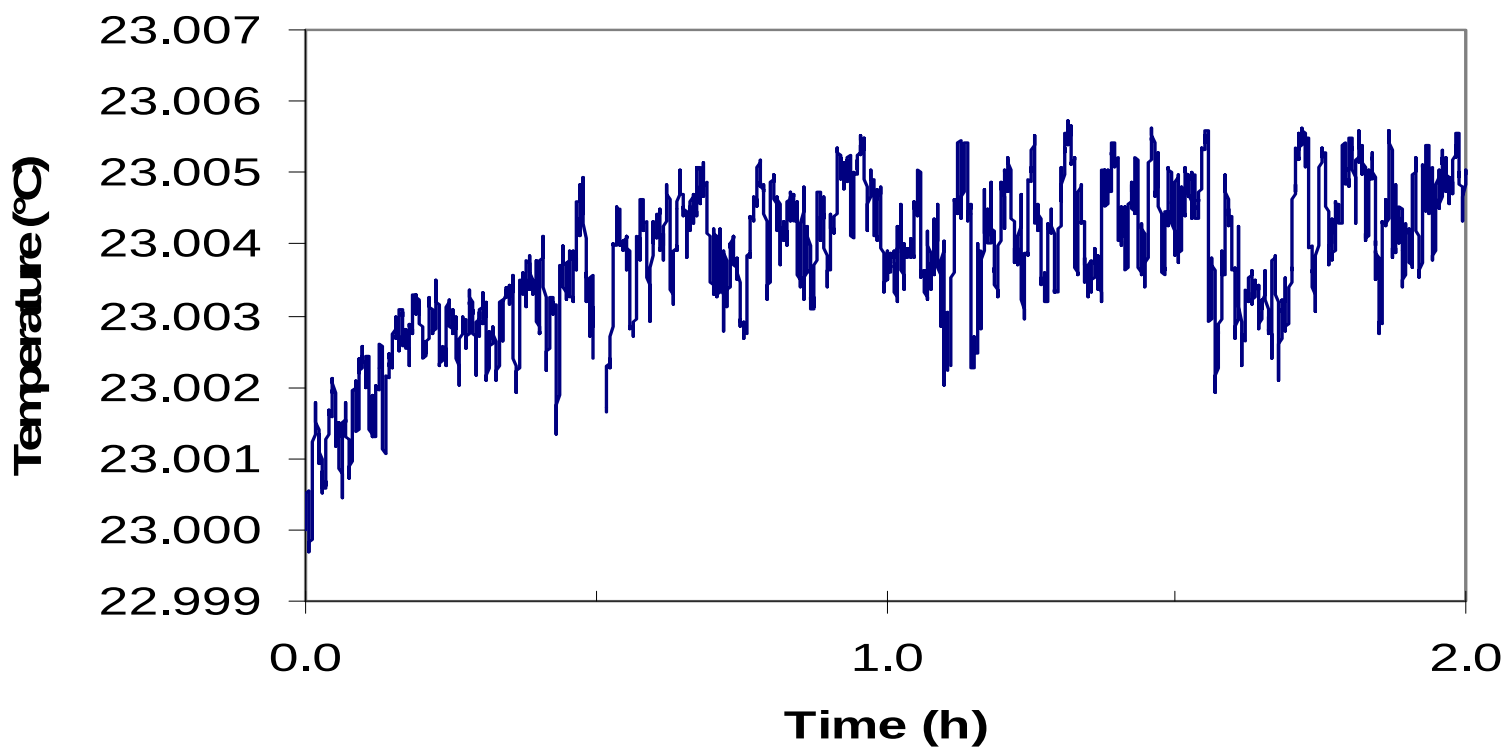

Fig. 4. Temperature stability in the box. Initial drift is due to a temperature set-point change.

## 4. EXPERIMENTAL RESULTS

### 4.1 1 Ω Standard

The time drift of the 1 Ω Standard is shown in Fig. 5. It shows a very high short-time stability and insensibility to temperature variations and thermal instabilities between its potentiometric connections. These measurements were made with a measurement method involving a high precision current comparator bridge [15]. The 2h spread (measurements standard deviation) was $4\times10^{-8}$ at the same level of high performance 1 Ω standard resistors in oil baths widely used in NMIs. The temperature dependence of the 1 Ω resistors net was evaluated from 22 °C to 24 °C changing the box temperature set point, resulting about $3\times10^{-6}$/°C.

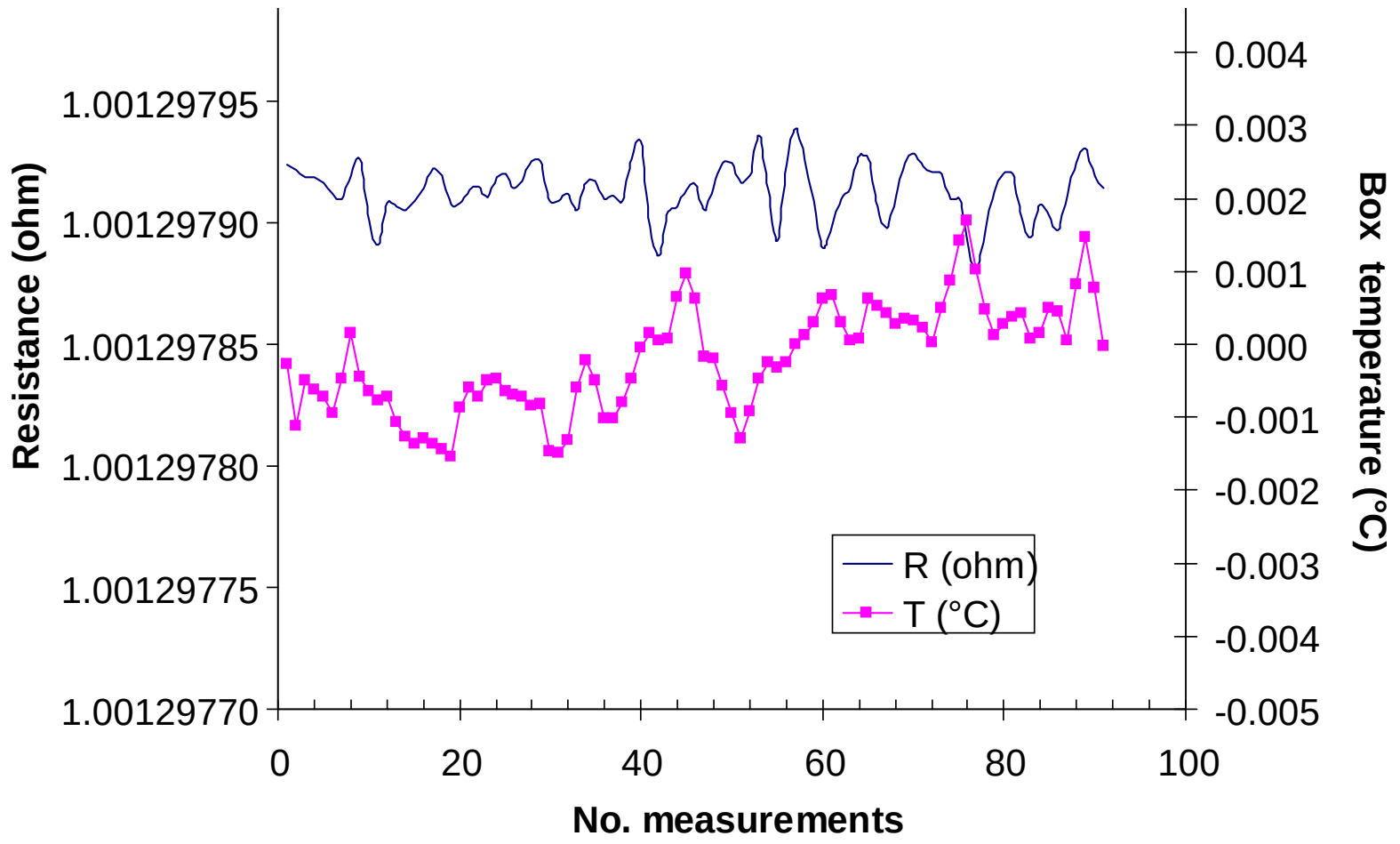

Fig. 5.Measurements on the 1 Ω Standard along with the box temperature drift with the temperature control set at 23 °C.

In addition in Fig. 6 it can be seen the ultra-high stability of the setup 1 Ω Standard in a typical calibration time at a single measurement current. Its measurements spread at 50 mA after stabilization was $1.3\times10^{-8}$ while the spreads, in the same conditions, of two oil-bath and one air top level commercial standard 1 Ω resistors were respectively $1.5\times10^{-8}$, $2.1\times10^{-7}$ and $2.1\times10^{-8}$. This test further confirms the benefit of the insertion of the 1 Ω Standard net in oil internally to the thermo-regulated box along with their internal potentiometric connections. This allows to reach a satisfactory stability during its calibration reducing time.

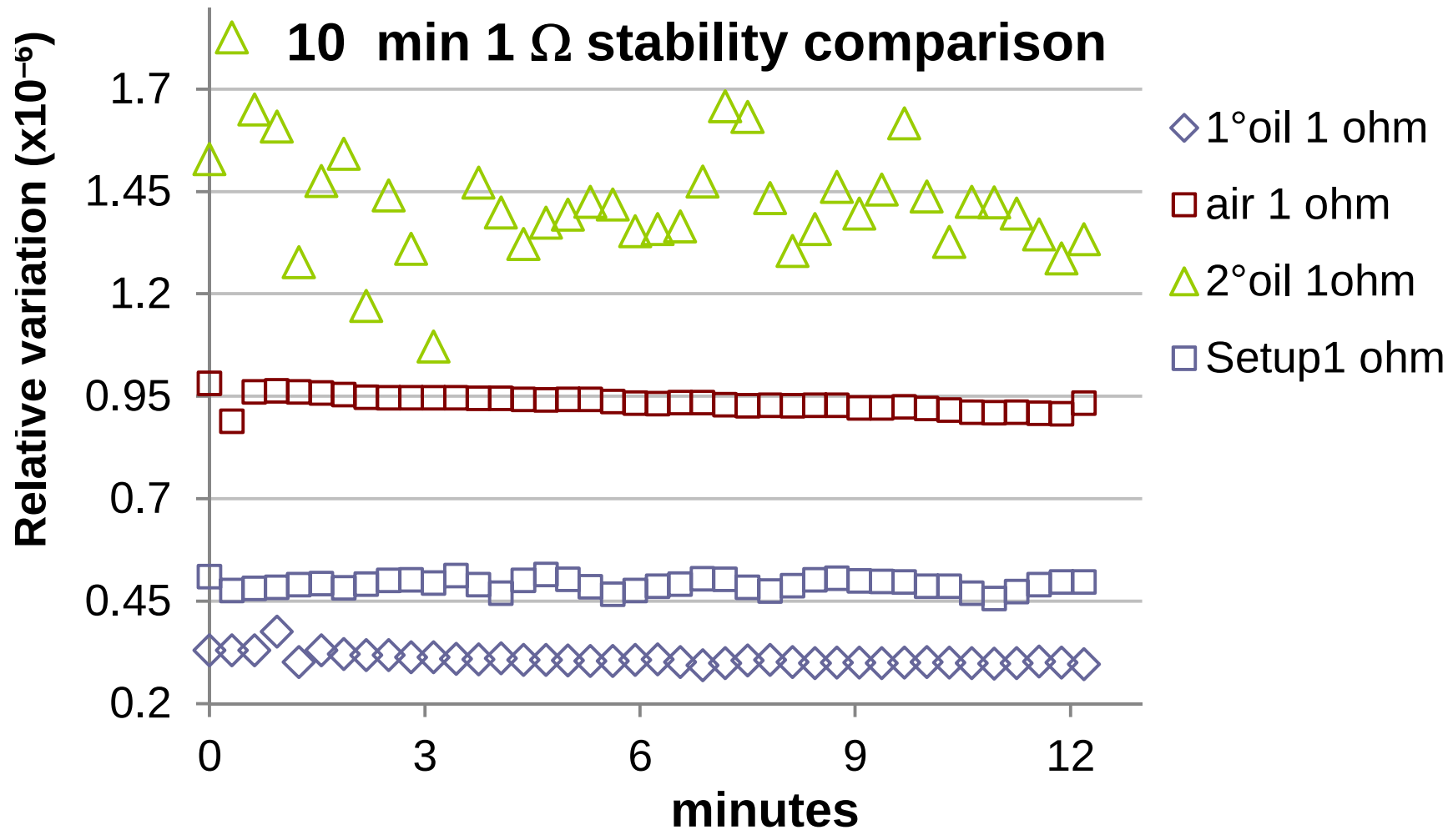

Fig. 6. Comparison of the behaviour of the setup 1 Ω Standard with three top level commercial 1 Ω Standards during a typical calibration time.

### 4.2 10 kΩ Standard

10 kΩ showed 2h a similar measurements spread ($5\times10^{-8}$) and temperature dependence of the net in the temperature range from 22 °C to 24 °C of $0.6\times10^{-6}$/°C although in an external ring of the box.

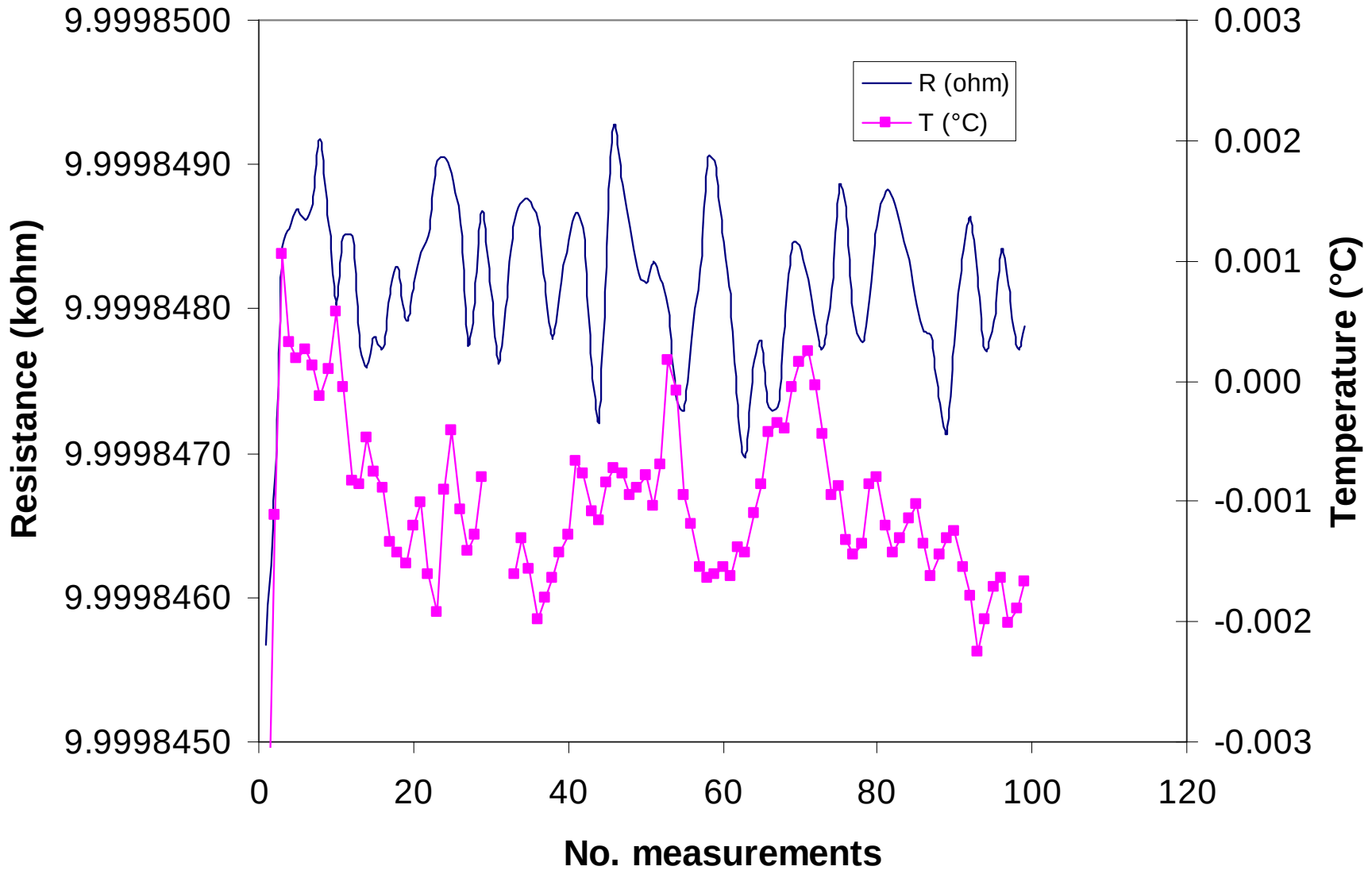

Fig. 7. Measurements on the 10 kΩ Standard with the temperature control set at 23 °C.

### 4.3 Week stability comparison among the setup Standards and top level commercial 1 Ω and 10 kΩ resistance Standards

A comparison of the week stability of the setup Standards and of the main top level commercial 1 Ω and 10 kΩ Resistance Standards was also carried on. This time period could be considered the mid period from their calibration at a NMI to their utilization to calibrate electrical instruments in customer Laboratories. In Fig. 8 the comparison of 1 Ω Standards is shown. The best stability was obtained by one commercial air Standard with a maximum value deviation during the week of $2.4\times10^{-8}$ while the commercial oil-bath and the setup Standards showed maximum value deviations respectively of $7.8\times10^{-8}$ and $4.3\times10^{-8}$. The spreads of the seven days values were $1.2\times10^{-8}$, $6.8\times10^{-8}$ and $2.5\times10^{-8}$ respectively for the air, oil and the setup Standards.

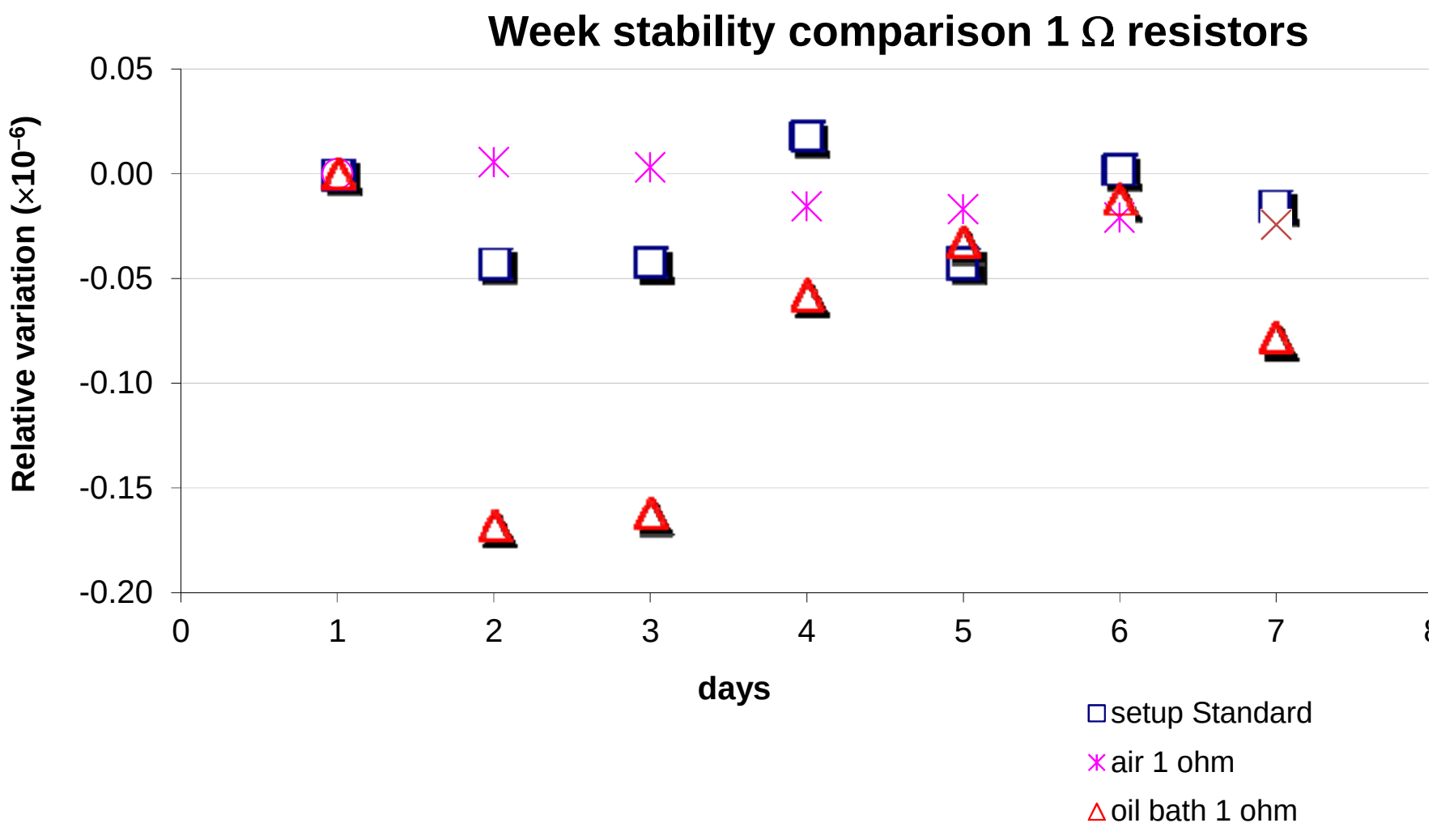

Fig. 8. Week drift of the 1 Ω setup Standard in comparison with two top level 1 Ω Standards, one of which in high stability oil bath.

In Fig. 9 the comparison of the 10 kΩ Standards is shown. The best stability was obtained by the setup Standard with a maximum value deviation during the week of $2.1\times10^{-8}$ while the other two commercial Standards showed maximum value deviations respectively of $4.3\times10^{-8}$ and $2.9\times10^{-8}$. The spreads of the seven days values were $0.7\times10^{-8}$, $1.8\times10^{-8}$ and $0.9\times10^{-8}$ respectively for the setup and the two commercial Standards.

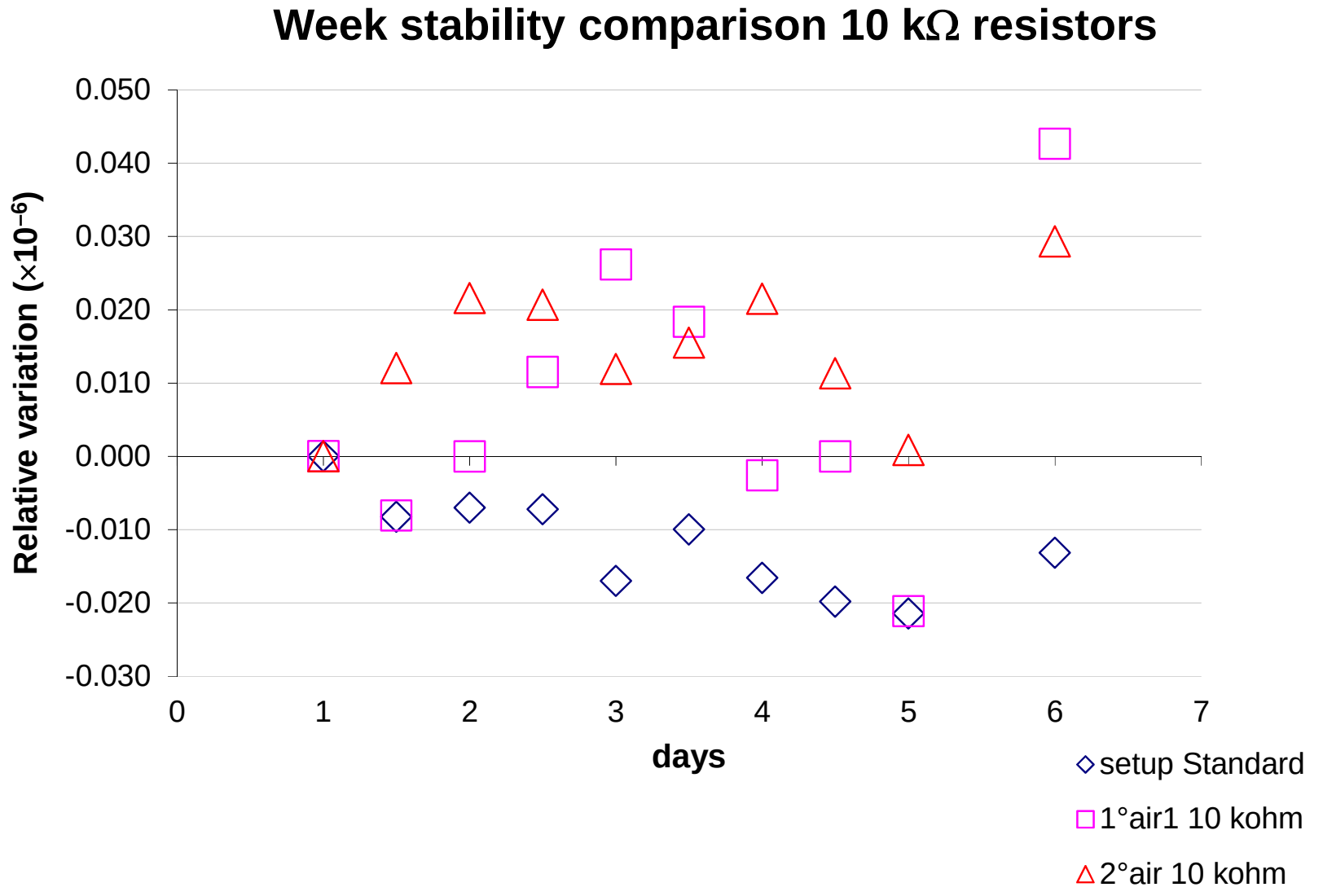

Fig. 9. Week drift of the 10 kΩ setup Standard in comparison with two top level 10 kΩ Standards.

.

**4.4 Mid-term stability and power coefficient of the 1 Ω and 10 kΩ Standards**

Fig. 10 shows the mid-term stability of the two Standards for about six months since the setup assembly.

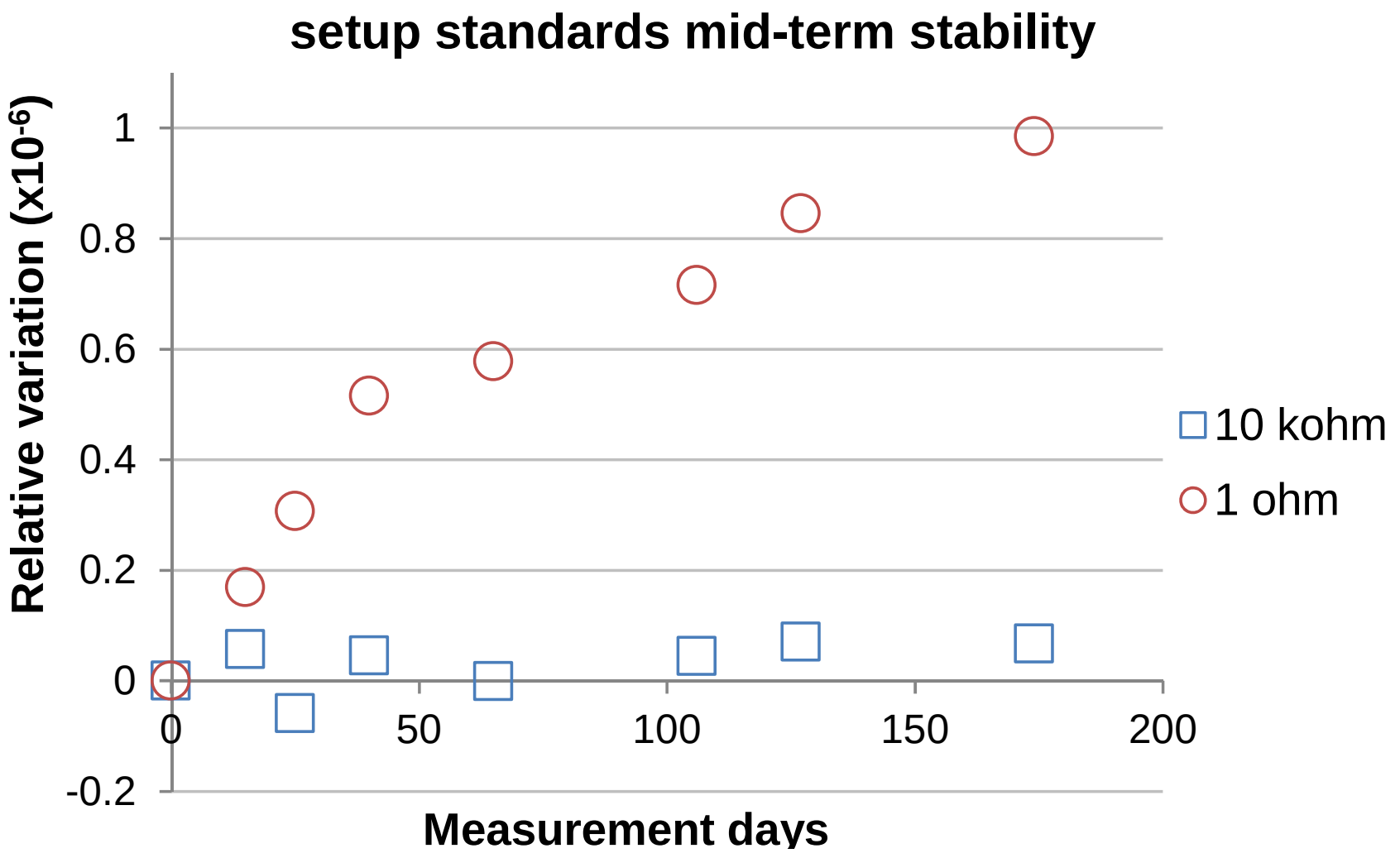

Fig. 10. Mid-term drift of the Standards measured since the setup assembly.

The 1 Ω showed an increasing .drift of $1.0\times10^{-6}$ while the 10 kΩ Standard showed a drift of $6.7\times10^{-8}$ as the resistors forming its net were stored for several years before the setup construction assuring to this Standard its better stability. This is of the same order of that an high accuracy commercial 10 kΩ resistor [16] and better than another [17]. The 1 Ω will be carefully monitored to verify if its value will reach a better stabilization, but already now its performance is on the order of the series [17]. The power coefficients of the two Standards were evaluated measuring them vs. high stability standard resistors with the same measurement system [15]. The results are reported in Table 1.

Table 1. Power coefficients of the setup Standards

| Standard | Power coefficient $\times 10^{-6}$/W) |
|---|---|
| 1 Ω | 1.7 |
| 10 kΩ | 2.7 |

The 1 Ω power coefficient allows to measure the Standard at currents up to 100 mA.

### 4.5 Temperature coefficients with the temperature control set at 23 °C.

To evaluate the temperature coefficients of the setup Standards with the temperature control set at 23°C and in the typical temperature conditions of electrical calibration laboratories, (23 ± 1) °C, the Standards were measured, after stabilization, at (22, 23 and 24) °C in a settable temperature laboratory. These temperature coefficients are reported in Table 2.

Table 2. Temperature coefficients of the setup Standards

| Standard | $\alpha_{23}$ ($\times 10^{-7}$ °K$^{-1}$) | β ($\times 10^{-7}$°K$^{-2}$) |
|---|---|---|
| 1 Ω | 5.5 | 1.0 |
| 10 kΩ | 0.6 | 1.4 |

The TCR of the 1 Ω Standard is not completely satisfactory presumably due to the TCR of the 10 Ω resistance elements involved in its net.

### 4.6 Transport effect.

The transport effect was evaluated transporting the setup turning off its temperature controller simulating the case of Standards belonging to external laboratories as the accredited calibration ones. The setup could be transported by car, van, plane and maintained for several hours or some days in not controlled temperature conditions till to the arrival to an external laboratory going beyond the battery capacity of its temperature control. For our test, the setup was transported in a suitable package by car with 2-3h travels, successively maintained in not-controlled temperature condition for at least 24h. Then the measurements were made in a thermo-regulated laboratory 24h after turning on again the temperature controller. Fig. 11 show the obtained results. The maximun deviations from the initial measurement before transports were $1.4\times10^{-7}$ and $1.7\times10^{-7}$ respectively for the 1 Ω and the 10 kΩ. The spreads of the obtained results were $5.6\times10^{-8}$ and $8.0\times10^{-8}$ respectively for the 1 Ω and the 10 kΩ.

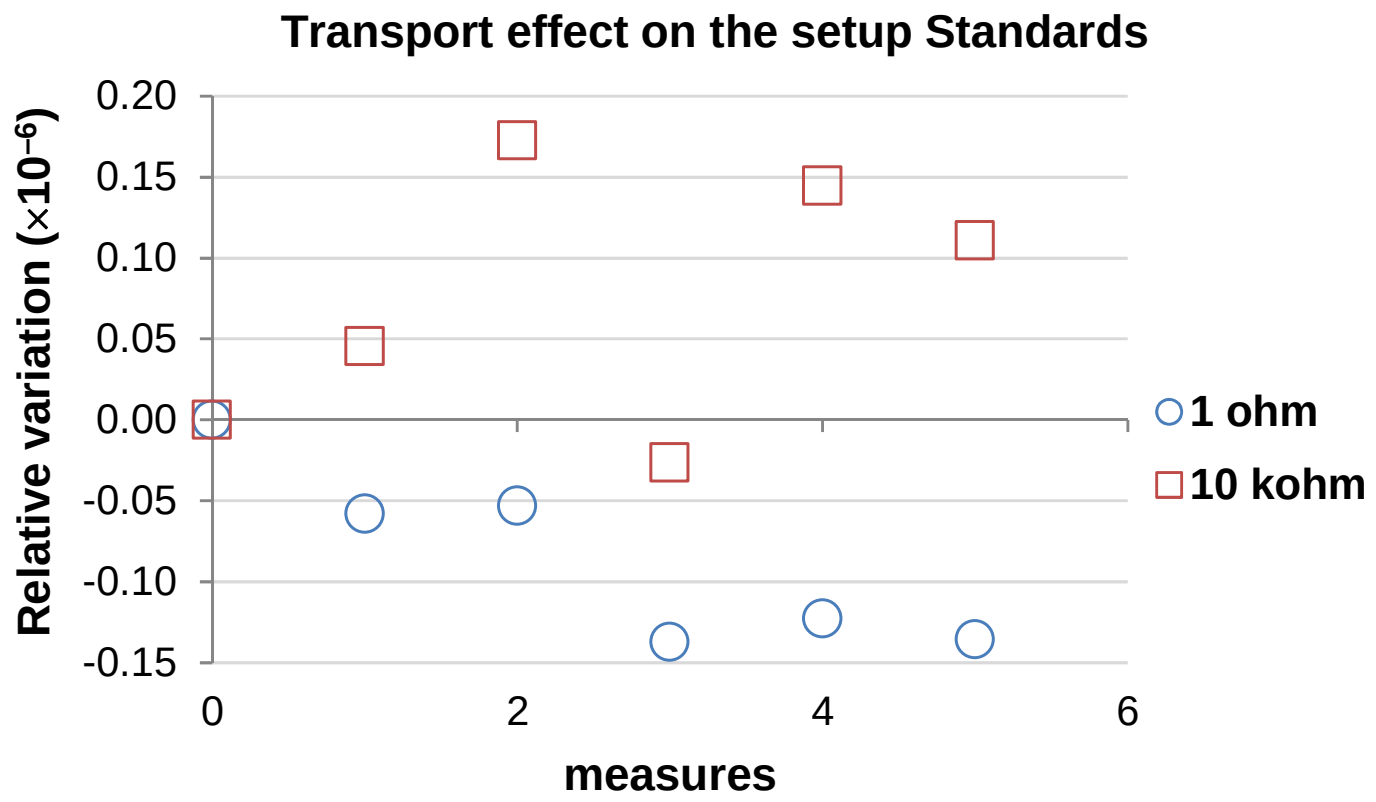

Fig. 11. Relative variation of the setup Standards measured after transports turning off the setup temperature control.

Both Standards were minimally affected by transport and by turning off the temperature control.

### 4.7 MFC's calibration and adjustment test.

The "Artifact calibration" is a process requiring only a small number of reference Standards with which high accuracy DMM's and MFC's can be calibrated and adjusted. At INRIM this operation for example on a MFC is performed in three steps [9]. With an initial verification a set of measurement points in which the MFC operates are compared with the reference system. After this, the adjustment is performed and then a final verification as in the first step checks the effectiveness of the adjustment. All the measurement deviations between the MFC and the reference system in the two verifications are recorded and inserted in the calibration certificates for customers. To check the suitability of the setup Standards to adjust electrical instruments, an initial verification of a high performance MFC was performed with the reference system utilized in its last calibration process observing that the measurement deviations in the 1 Ω and 10 kΩ points were unchanged with respect last final verification made some months before. Successively, an adjustment process with the setup Standards was performed. Then, a final verification as the in the first step to end the process confirmed the same deviations of the initial verification in the 1 Ω and 10 kΩ points. This result demonstrated that the adjustment with the setup Standards didn't introduce any systematic error in the adjustment process.

## 5. UNCERTANTY EVALUATIONS

### 5.1 Setup 1 Ω and 10 kΩ resistance Standards calibration and mid-term use uncertainties

The two setup Standards are calibrated vs. National Resistance Standard in the INRIM Resistance Calibration Laboratory by means of a measurement system involving high precision standard resistors put in a thermo-regulated oil-bath and a high performance current comparator bridge with expanded uncertainties of $1.7\times10^{-7}$ for the 1 Ω and $1.2\times10^{-7}$ for the 10 kΩ. With the data obtained in the setup Standards characterization in Tables 3 and 4 their mid-term use uncertainty budgets are given. It was assumed to use the setup Standards as local Standards initially for 180 days (mid-term period) without recalibration.

Table 3. 1 Ω mid-term use uncertainty.

| Source | type | 1σ ($\times10^{-7}$) |
|---|---|---|
| calibration | B | 0.85 |
| drift | B | 2.9 |
| emfs | B | 0.06[2] |
| Temperature | B | 3.2 |
| power | B | 0.02[3] |
| **Total RSS** | | **4.4** |

[2] This component was evaluated taking into account the maximum temperature difference (about 0.01 °C) between the resistors net and its internal connectors both maintained in oil.

[3] This component was evaluated considering the maximum possible applied power difference between the calibration at INRIM and in the utilization in a calibration Laboratory of the Standard.

Table 4. 10 kΩ mid-term use uncertainty.

| Source | type | 1σ (×$10^{-7}$) |
|---|---|---|
| calibration | B | 0.6 |
| drift | B | 0.2 |
| Temperature | B | 0.6 |
| power | B | $1.4^3$ |
| **Total RSS** | | **1.7** |

For a 95% confidence level the mid-term use uncertainties of the setup Standards are about $8.8\times10^{-7}$ and $3.4\times10^{-7}$ respectively for the 1 Ω and 10 kΩ.

**5.2 Use uncertainties for MFC's and DMM's calibration**

In the MFC's and DMM's calibration use uncertainty evaluation it can be considered a one week to one month-drift component as this calibration normally is performed after maximum a month since the calibration of the Standards at a NMI, but it is necessary to add a component due to the transport effect. The use uncertainties of the two setup Standards for DMM's and MFC's calibration are summarized in Table 5 and 6.

Table 5. 1 Ω use uncertainty for DMM's and MFC's calibration.

| Source | type | 1σ (×$10^{-7}$) |
|---|---|---|
| calibration | B | 0.85 |
| drift | B | 0.5 |
| emfs | B | 0.06 |
| Temperature | B | 3.2 |
| power | B | 0.02 |
| transport | B | 0.8 |
| **Total RSS** | | **3.4** |

Table 6. 10 kΩ use uncertainty for DMM's and MFC's calibration.

| Source | type | 1σ (×$10^{-7}$) |
|---|---|---|
| calibration | B | 0.6 |
| drift | B | 0.06 |
| Temperature | B | 0.6 |
| Power | B | 1.4 |
| transport | B | 1.0 |
| **Total RSS** | | **1.9** |

For a 95% confidence level the use uncertainties of the setup Standards for DMM's and MFC's calibration are $6.8\times10^{-7}$ and $3.8\times10^{-7}$ respectively for the 1 Ω and for the 10 kΩ.

**5.3 Uncertainties summary.**

In Table 7 a summary of the uncertainties at 2σ confidence level of the setup Standards is given.

Table 7. Setup Standards calibration, mid-term use, and for calibration of electrical instruments 2σ uncertainties.

| Standard | Calibration uncertainty | Mid-term use uncertainty | DMM-MFC calibration use uncertainty |
|---|---|---|---|
| 1 Ω | $1.7\times10^{-7}$ | $8.8\times10^{-7}$ | $6.8\times10^{-7}$ |
| 10 kΩ | $1.2\times10^{-7}$ | $3.4\times10^{-7}$ | $3.8\times10^{-7}$ |

These uncertainties meet those requested by DMM's and MFC's manufacturers to calibrate and adjust these instruments.

## 6. CONCLUSIONS

The characterization on the 1 Ω–10 kΩ setup Standards and a test to adjust a MFC gave satisfactory results as well as their use uncertainties, so the setup 1 Ω and 10 kΩ resistance Standards can be considered suitable for artifact calibration or as Reference Standards for maintaining the Resistance Unit in high level Laboratories. The cost of the development of the setup was of the same order of commercial top level 1 Ω and 10 kΩ standard Resistors as this it is a research prototype. Its cost could be significantly lower if its construction was carried out by a Resistance Standards manufacturer. Whit this setup it could be avoided the acquisition of oil-baths or the actual commercial top level highly expensive air resistors. Future aims will be the improvement of the temperature control to enhance the TCR of the 1 Ω Standard and the prosecution of the observation of its value, the evaluation the setup Standards humidity and pressure dependence to evaluate their attitude as traveling Standards for high level inter-laboratories comparisons.

## ACKNOWLEDGMENTS

The authors wish to thank the INRIM technician Marco Lanzillotti for his precious contribution in particular for the test of the calibration and adjustment of a MFC.